\documentclass[aps, prl,twocolumn, showpacs, superscriptaddress,preprintnumbers, amsmath, epsfig, floatfix,normalem]{revtex4}

\usepackage{graphicx}
\usepackage{dcolumn}
\usepackage{bm}
\usepackage{ulem}

\usepackage[utf8]{inputenc}
\usepackage[english]{babel}
\usepackage{keyval}
\graphicspath{{figs/}{figs:}{\figs}}
\setkeys{Gin}{width=0.90\columnwidth}

\usepackage[usenames,dvipsnames]{color}
\usepackage[]{ulem}

\newcommand{\comment}[1]{\textcolor{red}{#1}}
\renewcommand{\comment}[1]{\relax}

\newcommand{\todelete}[1]{\textcolor{green}{\sout{#1}}}
\renewcommand{\todelete}[1]{\relax}

\begin{document}
\title{Spectroscopic studies on the electronic and magnetic states of Co-doped perovskite manganite Pr$_{0.8}$Ca$_{0.2}$Mn$_{1-y}$Co$_{y}$O$_3$ thin films}
\date{\today}

\author{K. Yoshimatsu$^{*}$}
\affiliation{Department of Physics, University of Tokyo, 7-3-1 Hongo, Bunkyo-ku, Tokyo 113-0033, Japan}
\email{yoshimatsu@wyvern.phys.s.u-tokyo.ac.jp}

\author{H. Wadati}
\affiliation{Department of Applied Physics and Quantum-Phase Electronic Center (QPEC), University of Tokyo, 7-3-1 Hongo, Bunkyo-ku, Tokyo 113-8656, Japan}

\author{E. Sakai}
\affiliation{Photon Factory, Institute of Materials Structure Science, High Energy Accelerator Research Organization (KEK), 1-1 Oho, Tsukuba 305-0801, Japan}

\author{T. Harada}
\affiliation{Institute for Solid State Physics, University of Tokyo, 5-1-5, Kashiwanoha, Kashiwa, Chiba, 277-8581, Japan}

\author{Y. Takahashi}
\affiliation{Department of Physics, University of Tokyo, 7-3-1 Hongo, Bunkyo-ku, Tokyo 113-0033, Japan}

\author{T. Harano}
\affiliation{Department of Physics, University of Tokyo, 7-3-1 Hongo, Bunkyo-ku, Tokyo 113-0033, Japan}

\author{G. Shibata}
\affiliation{Department of Physics, University of Tokyo, 7-3-1 Hongo, Bunkyo-ku, Tokyo 113-0033, Japan}

\author{K. Ishigami}
\affiliation{Department of Physics, University of Tokyo, 7-3-1 Hongo, Bunkyo-ku, Tokyo 113-0033, Japan}

\author{T. Kadono}
\affiliation{Department of Physics, University of Tokyo, 7-3-1 Hongo, Bunkyo-ku, Tokyo 113-0033, Japan}

\author{T. Koide}
\affiliation{Photon Factory, Institute of Materials Structure Science, High Energy Accelerator Research Organization (KEK), 1-1 Oho, Tsukuba 305-0801, Japan}

\author{T. Sugiyama}
\affiliation{JASRI/SPring-8, Mikazuki-cho, Hyogo 679-5198, Japan}

\author{E. Ikenaga}
\affiliation{JASRI/SPring-8, Mikazuki-cho, Hyogo 679-5198, Japan}

\author{H. Kumigashira}
\affiliation{Photon Factory, Institute of Materials Structure Science, High Energy Accelerator Research Organization (KEK), 1-1 Oho, Tsukuba 305-0801, Japan}
\affiliation{PRESTO, Japan Science and Technology Agency, Kawaguchi, Saitama 332-0012, Japan}

\author{M. Lippmaa}
\affiliation{Institute for Solid State Physics, University of Tokyo, 5-1-5, Kashiwanoha, Kashiwa, Chiba, 277-8581, Japan}

\author{M. Oshima}
\affiliation{Department of Applied Chemistry, University of Tokyo, 7-3-1 Hongo, Bunkyo-ku, Tokyo 113-8656, Japan}
\affiliation{Synchrotron Radiation Research Organization, The University of Tokyo, Bunkyo-ku, Tokyo 113-8656, Japan}

\author{A. Fujimori}
\affiliation{Department of Physics, University of Tokyo, 7-3-1 Hongo, Bunkyo-ku, Tokyo 113-0033, Japan}

\begin{abstract}
We have investigated the electronic and magnetic properties of Co-doped Pr$_{0.8}$Ca$_{0.2}$MnO$_3$ thin films using various spectroscopic techniques. X-ray absorption and hard x-ray photoemission spectroscopy revealed that the substituted Co ions are in the divalent state, resulting in hole doping on the Mn atoms. Studies of element-selective magnetic properties by x-ray magnetic circular dichroism found a large orbital magnetic moment for the Co ions. These spectroscopic studies reveal that the substituted Co ions play several roles of hole doping for Mn, ferromagnetic superexchange coupling between the Co$^{2+}$ and Mn$^{4+}$ ions, and orbital magnetism of the Co$^{2+}$ ions. Competition among these complex interactions produces the unique electronic and magnetic behaviors including enhanced coercivity of the Co-doped Pr$_{0.8}$Ca$_{0.2}$MnO$_3$.
\end{abstract}
\maketitle
\section{I. Introduction}
\hspace*{3.35ex}Hole-doped perovskite manganites \textit{Re}$_{1-x}$\textit{A}$_{x}$MnO$_{3}$ (\textit{Re}: rare earth metal, \textit{A}: alkaline earth metal) have rich electronic and magnetic phase diagrams that depend on the hole concentration and the average ionic radius of the \textit{A}-site (\textit{Re}/\textit{A}) ions \cite{Ref1}. Antiferromagnetic insulating, ferromagnetic metallic, as well as charge- and orbital-ordered phases emerge for specific combinations of \textit{A}-site ionic radii and hole concentrations (\textit{e.g.}, Nd$_{0.5}$Sr$_{0.5}$MnO$_{3}$ and Pr$_{0.5}$Sr$_{0.5}$MnO$_{3}$) \cite{Ref1, Ref2}. Because dramatic phase transitions are induced in hole-doped perovskite manganites by applying external magnetic fields \cite{Ref2}, they have been intensively studied in order to utilize their colossal magnetoresistance and tunneling magnetoresistance (TMR) effects in spintronic devices \cite{Ref3}. For TMR devices, two ferromagnetic layers with different coercivities are necessary for independent reversal of magnetization directions under an applied magnetic field. However, perovskite manganites have the disadvantage of having quite small coercivities ($\sim$ 50 Oe at 5 K in La$_{1-x}$Sr$_{x}$MnO$_{3}$ (LSMO)) \cite{Ref3, Ref4}.  Nevertheless, using a ferromagnetic layer of Co-doped Pr$_{0.8}$Ca$_{0.2}$MnO$_3$ (PCMO), which has an enhanced coercivity, manganite-based spintronic devices with a TMR ratio greater than 120 \% were successfully fabricated recently \cite{Ref5}.  \\
\begin{figure}[htbp]
\begin{center}
\includegraphics[width=6cm]{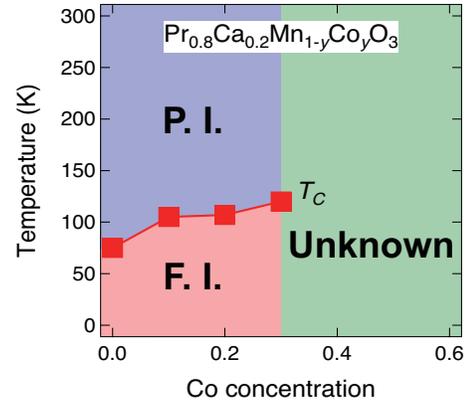}
\caption{(color online) Phase diagram of PCMCO (\textit{y} = 0--0.6) determined from the transport and SQUID measurements in Ref. 6. F. I. and P. I. denote the ferromagnetic insulating and paramagnetic insulating phases, respectively.}
\label{FIG1}
\end{center}
\end{figure}
 \hspace*{3.35ex}Figure 1 shows the phase diagram of Co-doped PCMO, Pr$_{0.8}$Ca$_{0.2}$Mn$_{1-y}$Co$_y$O$_3$ (PCMCO). PCMCO is a ferromagnetic insulator in the Co concentration range of \textit{y} = 0--0.3 \cite{Ref5, Ref6}. Its Curie temperature (\textit{T}$_{\rm C}$)  increases with Co substitution. While Co doping in PCMO is useful for improving the magnetic properties, the origin is not clear because of the lack of information about the electronic and magnetic states of Mn and Co. To solve this problem, we have employed x-ray absorption spectroscopy (XAS) and x-ray magnetic circular dichroism (XMCD) measurements, both of which offer element-selective capabilities. From the XAS measurements, the valence states of the two transition metal ions could be determined and from the XMCD measurements, the element-selective magnetic information could be obtained. In addition to these two spectroscopic methods, hard x-ray photoemission spectroscopy (HAXPES) measurements were performed to obtain information about the electronic structure near the Fermi level (\textit{E}$_{\rm F}$) that dominates the transport properties. \\
\section{II. Experimental}
\hspace*{3.35ex}Epitaxial PCMCO (\textit{y} = 0--0.3) thin films were grown onto Nb-doped SrTiO$_3$ (001) substrates by  pulsed laser deposition. Detailed growth conditions of PCMCO films (grown on LSAT substrates) were described elsewhere. \cite{Ref6} In the present study, use of the Nb-SrTiO$_3$ substrates was necessary for preventing charging effects in the spectroscopic experiments. All PCMCO films were coherently grown and were therefore under compressive strain from the substrates, as shown by four-circle x-ray diffraction measurements (not shown). The \textit{T}$_{\rm C}$, determined from the \textit{M}-\textit{T} curves (not shown), ranged from 75 (\textit{y} = 0) to 120 K (\textit{y} = 0.3) \cite{Ref6, Ref7}. The saturation magnetizations were determined from the \textit{M}-\textit{H} curves.\cite{Ref8}\\
\hspace*{3.35ex}The XMCD measurements were performed at the undulator beamline BL-16A of the Photon Factory, using an XMCD system equipped with a vector-type superconducting magnet \cite{Ref9}. A magnetic field of \textit{B} = 1 T was applied parallel to the photon direction, and the angle between the sample surface and the magnetic field was set to 30$^\circ$. The photon helicity was reversed and the magnetic field direction was fixed for the XMCD measurements. The degree of circular polarization was $\pm$95$\pm$4 \% \cite{Ref10}. All  measurements were performed at 20 K, which is well below the \textit{T}$_{\rm C}$ of PCMCO. The XAS spectra are defined as the average of the absorption spectra taken with positive and negative photon helicities.\\
\hspace*{3.35ex}The HAXPES measurements were carried out at the undulator beamline BL47XU of SPring-8. Synchrotron radiation from the undulator was monochromatized to 7.94 keV with a Si 111 double crystal monochromator and a Si 444 channel-cut monochromator, which reduced the energy bandwidth to 80 meV. All HAXPES spectra were recorded at room temperature using a Scienta R-4000 electron energy analyzer with a total energy resolution of 250 meV. The \textit{E}$_{\rm F}$ of the samples was referenced to that of a gold foil that was in electrical contact with the samples.\\
\section{III. Results and Discussion}
\begin{figure}[htbp]
\begin{center}
\includegraphics[width=7cm]{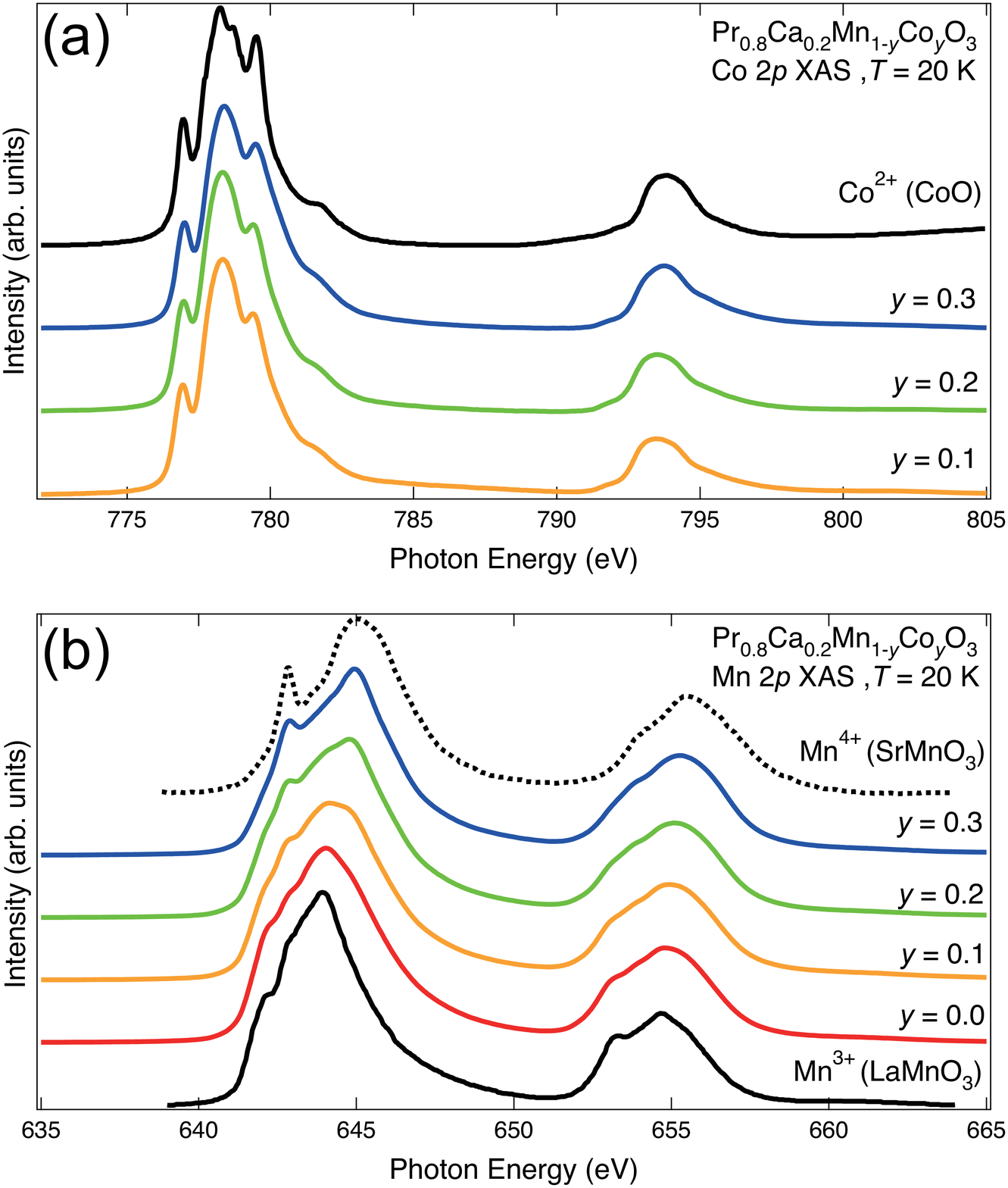}
\caption{(color online) XAS spectra of PCMCO (\textit{y} = 0--0.3). (a) Co 2\textit{p} XAS spectra. The XAS spectrum of Co$^{2+}$ for CoO is also shown as a reference. (b) Mn 2\textit{p} XAS spectra. The XAS spectra of Mn$^{3+}$ (LaMnO$_3$) and Mn$^{4+}$ (SrMnO$_3$) are also shown as references.}
\label{FIG2}
\end{center}
\end{figure}
\hspace*{3.35ex}Figure 2(a) shows the Co 2\textit{p} XAS spectra of PCMCO (\textit{y} = 0.1--0.3). For comparison, the XAS spectrum of CoO is also included \cite{Ref11}. Each Co 2\textit{p} XAS spectrum shows two features of the \textit{L}$_3$ (h$\nu$ = 775--785 eV) and \textit{L}$_2$ (h$\nu$ = 790--800 eV) edges. The Co \textit{L}$_3$ edge consists of three sharp peak structures located at 777, 778, and 779.5 eV. The \textit{L}$_3$ edge also has shoulder structures on the higher photon energy side ($\sim$782 eV). On the other hand, the \textit{L}$_2$ edge has a broad structure centered at 793.5 eV. Because the XAS spectral shapes strongly depend on the valence state, the valence of the Co ions can be evaluated in comparison with reference XAS spectra: The Co 2\textit{p} XAS spectra of PCMCO are quite similar to the spectrum of CoO, suggesting that the Co ions in PCMCO and CoO have an identical valence state, \textit{i.e.}, the divalent Co ions in CoO$_6$ octahedra have the high-spin configuration ($t_{2g}^5$, $e_{g}^2$) \cite{Ref11}. Note that the octahedrally-coordinated trivalent Co ions in LaCoO$_3$ show a completely different XAS spectrum (not shown) \cite{Ref11}. These results indicate that the Co ions in PCMCO are in the divalent state with a high-spin configuration in PCMCO, regardless of the Co concentration (\textit{y} = 0.1--0.3).\\	
\begin{figure}[htbp]
\begin{center}
\includegraphics[width=7cm]{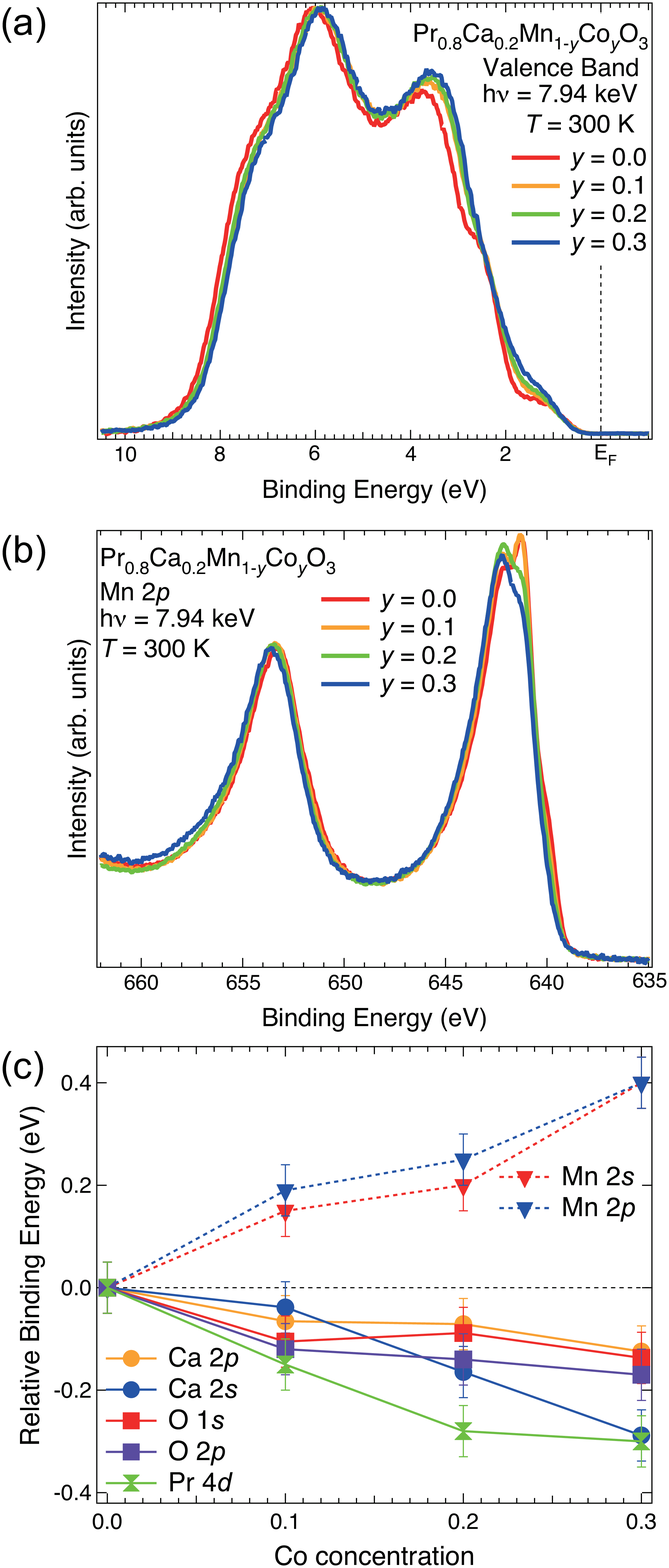}
\caption{(color online)HAXPES spectra of PCMCO (\textit{y} = 0--0.3). (a) Valence-band spectra. (b) Mn 2\textit{p} core level spectra. (c) Plot of the core-level shifts as a function of the Co concentration.}
\label{FIG3}
\end{center}
\end{figure}
\hspace*{3.35ex}Because the divalent Co ions substitute for the Mn sites with valences between 3+ and 4+, holes should be doped into the remaining Mn atoms. To evaluate the valence of the Mn ions, the Mn 2\textit{p} XAS spectra were measured as shown in Fig. 2(b). For reference, the Mn 2\textit{p} XAS spectra of Mn$^{3+}$ (LaMnO$_3$) \cite{Ref11} and Mn$^{4+}$ (SrMnO$_3$) \cite{Ref11} are also shown in the same figure. For the parent PCMO compound, the nominal valence of the Mn ions is +3.2 and the XAS spectral shape is similar to that of LaMnO$_3$. The XAS spectrum of PCMCO (\textit{y} = 0.3) is somewhat different from that of LaMnO$_3$, especially in the region of the characteristic peak structure of the \textit{L}$_3$ edge. This peak is also observed in the XAS spectrum of SrMnO$_3$, suggesting that the valence of the Mn ion increases with Co substitution. In addition, the center of gravity of the Mn \textit{L}$_3$ peak is shifted toward higher photon energies with increasing Co concentration. Because the same shift has been reported in the hole-doped perovskite manganite LSMO \cite{Ref12}, the present result can be taken as evidence for hole doping of the Mn atoms induced by the Co substitution. \\
\hspace*{3.35ex}The Co substitution in PCMO also influences the valence band and core levels measured by HAXPES as shown in Fig. 3.  Figure 3 (a) shows the valence-band spectra of PCMCO (\textit{y} = 0--0.3). The valence band mainly consists of two prominent O 2\textit{p}-derived structures centered at 6 and 3.5 eV below \textit{E}$_{\rm F}$. The deeper structure consists of Mn 3\textit{d}-O 2\textit{p} bonding states and the shallower one of O 2\textit{p} non-bonding states \cite{Ref13}. These states are shifted toward lower binding energies with Co substitution, suggesting that a downward chemical potential shift occurs due to the hole doping. Pr 4\textit{d}, Mn 3\textit{d} \textit{t}$_{2g}$, and Co 3\textit{d} \textit{t}$_{2g}$ states are expected to exist around the shoulder of the O 2\textit{p} non-bonding states \cite{Ref13} although they cannot be isolated in the spectra. The density of states (DOS) closest to \textit{E}$_{\rm F}$ does not reach \textit{E}$_{\rm F}$, in agreement with the insulating nature of PCMCO. With increasing Co concentration, the DOS near \textit{E}$_{\rm F}$ increases. When the hole doping on Mn atoms occurs, the Mn 3\textit{d} \textit{e}$_{g}$ DOS should be reduced because the doped holes primarily enter the Mn 3\textit{d} \textit{e}$_{g}$ band. Therefore, we conclude that the DOS is derived from both Mn 3\textit{d} \textit{e}$_{g}$ and Co 3\textit{d} \textit{e}$_{g}$ states and that they appear nearly at the same energies.\\
\begin{figure}[htbp]
\begin{center}
\includegraphics[width=7cm]{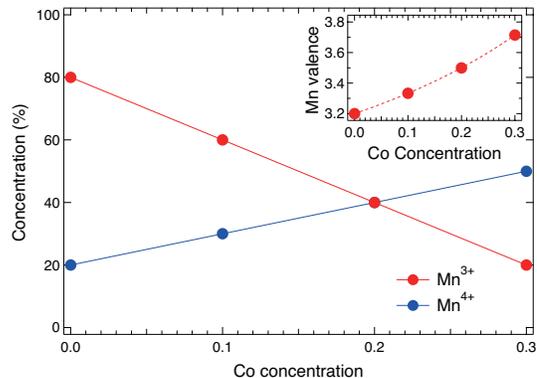}
\caption{(color online) Proportions of Mn ions with different valences as a function of the Co concentration in PCMCO. The inset shows the nominal Mn valence as a function of Co substitution. }
\label{FIG4}
\end{center}
\end{figure}
\hspace*{3.35ex}The Mn 2\textit{p} core-level photoemission spectra show evolution of the spectral line shape with Co substitution. Figure 3(b) shows the Mn 2\textit{p} core-level spectra of PCMCO (\textit{y} = 0--0.3), the Mn 2\textit{p}$_{3/2}$ peak shows higher intensity on the lower binding energy side for lower Co concentrations (\textit{y} = 0 and 0.1) while for higher Co concentrations (\textit{y} = 0.2 and 0.3), Mn 2\textit{p}$_{3/2}$ peak shows higher intensity on the higher binding energy side. The chemical shifts of the Mn core levels also support the Mn hole doping scenario. In contrast to the Mn 2\textit{p}$_{3/2}$ peaks, the Mn 2\textit{p}$_{1/2}$ peak structure remained unchanged, but the peak position shifted toward higher binding energies with increasing Co concentration. This energy shift is opposite to that of the O 2\textit{p} states as shown in Fig. 3(a). To investigate this behavior, we have plotted the energy shifts of the other core levels as a function of Co concentration in Fig. 3(c). All core levels except for the Mn-derived states are shifted toward lower binding energies by about 0.2 eV. Similar core-level shifts have been reported in the hole doped PCMO \cite{Ref13}. The Mn 2\textit{p} and Mn 2s core levels show opposite energy shifts to the O, Pr, and Ca core levels. \\
\hspace*{3.35ex}The shift of a core level with varying chemical composition  is given by,
\begin{eqnarray}
\Delta E &=& \Delta \mu +K\Delta Q +\Delta V_M-\Delta E_R
\end{eqnarray} 
where $\Delta \mu$ is the change in the chemical potential, $\Delta Q$ is the change in the number of valence electrons on the atom considered, $\Delta V_M$ is the change in the Madelung potential, and $\Delta E_R$ is the change in the extra-atomic relaxation energy derived from changes in the screening of the core-hole potential by metallic conduction electrons and surrounding ions \cite{Ref14}. For the Pr, Ca and O atoms, $\Delta \mu$ dominates the core-level shifts because the valence of Pr, Ca and O ions is independent of Co doping ($\Delta Q$ = 0). $\Delta E_R$ is negligible because PCMCO is  an insulator. A change of the Madelung potential ($\Delta V_M$) can also be excluded because it would cause shifts of the core levels of the O anion and the Pr and Ca cations in different directions. Meanwhile, the Mn core-level shifts are mainly influenced by a change in the number of valence electrons, which is called the chemical shift, although a change in the chemical potential is also included \cite{Ref15}. The Co substitution increases the Mn valence due to hole doping, which corresponds to a smaller number of electrons ($\Delta Q <$ 0). The direction of the core-level shift for the transition metal atoms is determined by a competition between the chemical shift and the chemical potential shift \cite{Ref16}. Such opposite core-level shifts between elements have been observed in many complex transition-metal oxides, including the high-\textit{T}$_{\rm C}$ cuprate La$_{2-x}$Sr$_{x}$CuO$_{4}$ \cite{Ref17}. The present core-level shifts are fully consistent with hole-doping at the Mn sites, justifying our basic picture.\\
\hspace*{3.35ex}Assuming that all  Co ions in PCMCO (\textit{y} $\leq$ 0.3) are divalent, the valence of Mn ions can be estimated to be (3.2-2\textit{y})/(1-\textit{y}). The resultant Mn valence as a function of Co concentration is plotted in the inset of Fig 4. The nominal valence of Mn monotonically increases with increasing Co concentration. From the nominal Mn valence, the concentrations of Mn$^{3+}$ and Mn$^{4+}$ ions can be deduced and are plotted in Fig. 4.  The ratio of Mn$^{4+}$ linearly increases and that of Mn$^{3+}$ linearly decreases as a function of Co concentration. These results are utilized to estimate the total magnetization and the orbital magnetic moment of each element using the XMCD sum rules \cite{Ref18}. \\
\hspace*{3.35ex}Figure 5(a) shows the Co 2\textit{p} XMCD spectra of PCMCO. The XMCD spectrum of La$_2$MnCoO$_6$ (LMCO) is also plotted as a reference \cite{Ref11}. All the XMCD spectra of PCMCO are normalized to the integrated XAS intensity as shown in Fig. 2(a). The spectral line shapes of PCMCO are quite similar to those of LMCO. In addition, the XMCD spectra of PCMCO (\textit{y} = 0.1--0.3) are similar to each other in terms of line shapes and intensities, indicating that the same degree of ferromagnetic Co$^{2+}$ states are present in PCMCO, regardless of the Co concentration. The stronger negative and weaker positive XMCD signals at the \textit{L}$_3$ and \textit{L}$_2$ edges represent a finite orbital magnetic moment parallel to the spin magnetic moment. A quantitative analysis is presented below.\\
\begin{figure}[htbp]
\begin{center}
\includegraphics[width=7cm]{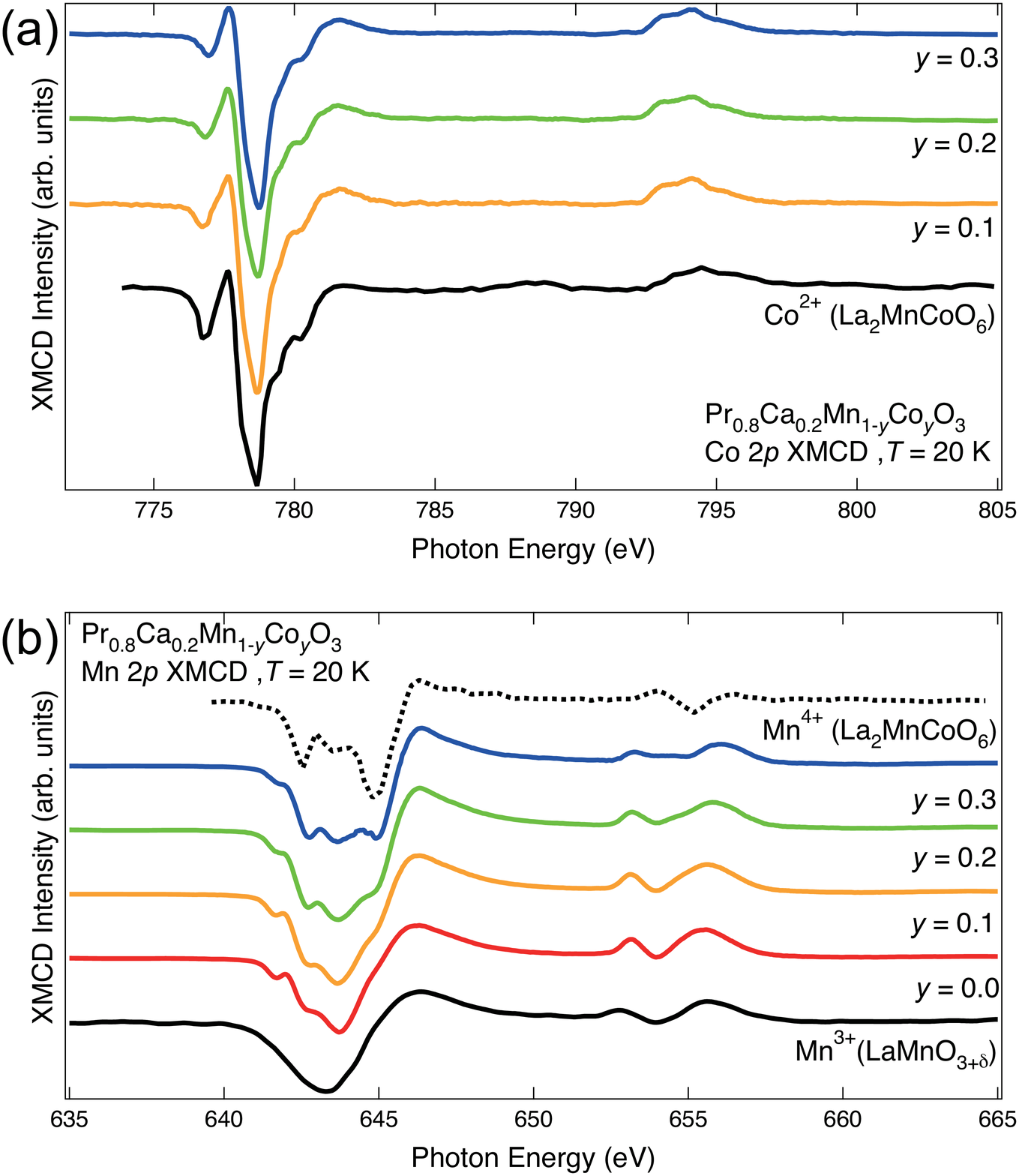}
\caption{(color online) XMCD spectra of PCMCO (\textit{y} = 0.0--0.3). (a) Co 2\textit{p} XMCD spectra. The XMCD spectrum of Co$^{2+}$ for La$_2$MnCoO$_6$ (LMCO) is also shown for reference. (b) Mn 2\textit{p} XMCD spectra. The XMCD spectra of Mn$^{3+}$ (LaMnO$_{3+\delta}$) and Mn$^{4+}$(LMCO) are also shown as references.}
\label{FIG5}
\end{center}
\end{figure}
\hspace*{3.35ex}Mn 2\textit{p} XMCD spectra were also recorded as shown in Fig. 5(b). The spectra have also been normalized to the integrated XAS intensity shown in Fig. 2(b). The reference XMCD spectra of Mn$^{3+}$ (LaMnO$_{3+\delta}$) and Mn$^{4+}$ (LMCO) are shown in the same figure. The XMCD signal is negative at the \textit{L}$_3$ edge and positive at the \textit{L}$_2$ edge, which indicates that the spin magnetic moment of the Mn ions is aligned parallel to that of the Co ions. This result indicates that ferromagnetic coupling exists between the Mn and Co ions. \\
\hspace*{3.35ex}In contrast to the Co 2\textit{p} XMCD spectra, the Mn 2\textit{p} XMCD spectral line shape changes systematically with Co concentration. The XMCD spectrum of the parent compound PCMO (with the nominal valence of Mn$^{3.2+}$) is similar to that of the oxygen-excess LMO (Mn$^{(3+2\delta )+})$ \cite{Ref19}. With increasing Co concentration, a kink structure appears at about 645 eV of the \textit{L}$_3$ edge in the XMCD spectra of PCMCO (\textit{y} = 0.1 and 0.2). It finally becomes a peak in the \textit{y} = 0.3 sample. From comparison with the reference XMCD spectrum of Mn$^{4+}$ (LMCO), we conclude that the peak comes from the ferromagnetic Mn$^{4+}$ ion.  \\
\hspace*{3.35ex}For element-selective quantitative analysis of the ferromagnetic moment in PCMCO, the XMCD spin and orbital sum rules are applied to the Mn and Co 2\textit{p} XMCD spectra. The spin and orbital magnetic moments are obtained using the following equations,
\begin{eqnarray}
\nonumber  m_{\rm spin}+7m_{T_{z}} &=& \frac{-2(\Delta I_{L3}-2\Delta I_{L2})}{I_{L3}+I_{L2}}n_{h}\mu_{\rm B},\\
m_{\rm orbital}&=&\frac{-4(\Delta I_{L3}+\Delta I_{L2})}{3(I_{L3}+I_{L2})}n_{h}\mu_{\rm B}
\end{eqnarray} 
where $I_{Li}$ and $\Delta I_{Li}$ represent the integrals of the XAS and XMCD spectra in the \textit{L}$_i$ edge region, \textit{n}$_{h}$ is the number of holes per atom and $\mu _{\rm B}$ is the Bohr magneton. Here, $m_{T_{z}}$ denotes the magnetic dipole moment, which is negligible compared to \textit{m}$_{\rm spin}$ for ions in the cubic symmetry \cite{Ref20}.  For XMCD sum rule analysis, the number of holes on each atom needs to be known. In this analysis, the number of holes on the Co atoms is assumed from the nominal valence: the Co$^{2+}$ (\textit{d}$^7$) ion has 3 holes in the whole Co concentration range because the Co 2\textit{p} XAS and XMCD spectra of Co$^{2+}$ are well reproduced by the cluster model calculation of Co$^{2+}$ ion without charge transfer from the oxygen ligands. The number of holes on the Mn atoms is assumed using the following equation with linear interpolation in order to take the charge transfer into account.\\
\begin{eqnarray}
\nonumber n_{h, \rm Mn 3\textit{d}}=(4+\Delta n_{\rm Mn 3\textit{d}})\times (4-z)+\\
(3+\Delta n^{'}_{\rm Mn 3\textit{d}})\times (3-z) 
\end{eqnarray} 
where the parameters are set to $\Delta$n$_{Mn 3d}$ = 0.5 for LaMnO$_3$ and $\Delta$n$^{'}_{Mn 3d}$ = 0.8 for SrMnO$_3$ \cite{Ref19} and \textit{z} is the nominal valence of the Mn ions, estimated from Fig. 4. The spin and orbital magnetic moments estimated in this way for Co and Mn are shown in Fig. 6. \\
\hspace*{3.35ex}Figure 6 shows the magnetic moments calculated using the XMCD sum rules as a function of Co concentration. The ferromagnetic moment of the Co ion is roughly constant with Co substitution. The spin and orbital magnetic moments of Co atoms (\textit{m}$_{\rm Co, spin}$ and \textit{m}$_{\rm Co, orbital}$) are about 1 and 0.25 $\mu _{\rm B}$/Co, respectively. The sum ($\sim$1.25 $\mu _{\rm B}$/Co) is smaller than the total magnetization obtained from SQUID measurements (\textit{M}$_{\rm total, SQUID}$ = 1.5$\sim$2.0 $\mu _{\rm B}$/unit cell) \cite{Ref8}, suggesting that the magnetization of Mn atoms increases with Co doping. In fact, the spin magnetic moment of the Mn atoms (\textit{m}$_{\rm Mn, spin}$) increases with Co doping. On the other hand, the orbital magnetic moment of Mn atoms (\textit{m}$_{\rm Mn, orbital}$) remains almost zero in the whole Co concentration range. A very small orbital magnetic moment has been reported for several perovskite Mn oxides, (\textit{e.g.} LSMO) \cite{Ref19}. The total magnetization calculated using the XMCD sum rules (\textit{M}$_{\rm total, XMCD}$) was obtained from the following equation:
\begin{eqnarray}
\nonumber M_{\rm total, XMCD}=(1-y)\times(m_{\rm Mn, spin}+m_{\rm Mn, orbital})+\\
y\times (m_{\rm Co, spin}+m_{\rm Co, orbital})
\end{eqnarray}
which yielded an \textit{M}$_{\rm total, XMCD}$ of about 1.5 $\mu _{\rm B}$/unit cell for the whole Co concentration range, consistent with \textit{M}$_{\rm total, SQUID}$. \\
\hspace*{3.35ex}Finally, we briefly discuss the origin of the interesting magnetic behavior of PCMCO.  The parent compound, PCMO has a nominal Mn valence of 3.2+ with high spin configuration, suggesting a maximum spin magnetic moment of 3.8 $\mu _{\rm B}$/unit cell. The experimental magnetic moment obtained from the XMCD and SQUID measurements (1.5$\sim$2.0 $\mu _{\rm B}$/unit cell) are much smaller than the ideal value. This suggests that antiferromagnetic coupling, \textit{e.g.}, due to a superexchange interaction between neighboring Mn$^{3+}$ ions via oxygen ions affects the magnetic properties of PCMO but is weakened by hole doping. \\
\begin{figure}[htbp]
\begin{center}
\includegraphics[width=8cm]{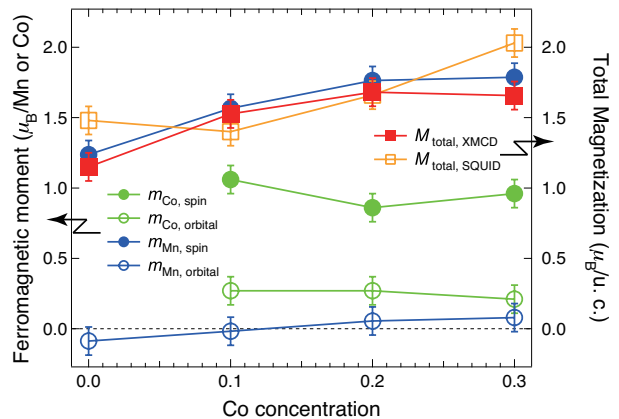}
\caption{(color online) Ferromagnetic moments in PCMCO as a function of Co concentration. The total magnetization deduced from XMCD (\textit{M}$_{\rm total, XMCD}$) has been obtained using the \textit{m}$_{\rm spin}$ and \textit{m}$_{\rm orbital}$ of Mn and Co atoms and the B-atom concentration shown in Fig. 4. The total magnetization deduced from SQUID measurements (\textit{M}$_{\rm total, SQUID}$) has been determined from the \textit{M}-\textit{H} curves \cite{Ref8}.}
\label{FIG6}
\end{center}
\end{figure}
\hspace*{3.35ex}Considering the phase diagram of bulk PCMO \cite{Ref21}, the ferromagnetic ground states appears only in a Ca concentration range from \textit{x} = 0.15 (Mn$^{3.15+}$) to 0.3 (Mn$^{3.3+}$). However, the Mn atoms have a nominal valence of 3.5+ and 3.7+ in PCMCO with \textit{y} = 0.2 and \textit{y} = 0.3, respectively, suggesting that Co ions play other roles besides hole doping, such as including a ferromagnetic superexchange interaction between Co$^{2+}$ and Mn$^{4+}$, as expected from Goodenough-Kanamori rule \cite{Ref22, Ref23}. Because of electrostatic interactions, Mn atoms neighboring the Co$^{2+}$ ions tend to be tetravalent, and these two ions are expected to be ferromagnetically coupled with each other in the whole Co concentration range. As a result, the magnetization of Co is almost independent of the Co concentration while the magnetization of Mn increases with Co concentration through a double-exchange interaction enhanced by the hole doping. The combination of ferromagnetic superexchange and double-exchange interactions aligns the spins of Co$^{2+}$ and Mn$^{3+/4+}$ ions, stabilizing the ferromagnetic ground state in this system and leading to an increase of the magnetization of Mn and \textit{T}$_{\rm C}$. \\
\hspace*{3.35ex}Another characteristic magnetic behavior of PCMCO is the enhanced coercivity compared to PCMO \cite{Ref6}. The substituted Co ions have a larger orbital magnetic moment than Mn, which induces the larger coercivity for the higher Co concentrations. \\
\section{IV. Conclusion}
\hspace*{3.35ex}In summary, we have investigated the electronic and magnetic properties of Co-doped PCMO using HAXPES, XAS, and XMCD. The HAXPES and XAS results reveal the valence states of the Co and Mn ions. The Co ions are in a high-spin divalent state in the whole Co concentration range, resulting in hole doping at the Mn sites. The XMCD results indicate that the Co and Mn spins are ferromagnetically coupled, which we attribute to the ferromagnetic superexchange interaction between the Co$^{2+}$ and Mn$^{4+}$ ions in addition to the double-exchange interaction between the Mn$^{3+}$ and Mn$^{4+}$ ions. The Co$^{2+}$ ions with large orbital magnetic moments are expected to induce the large coercivity in PCMCO. \\
\section{Acknowledgement}
This work was supported by a Grant-in-Aid for Scientific Research (S22224005) andResearch Activity Start-up (25887021) from the Japan Society for the Promotion of Science (JSPS) program and the Quantum Beam Technology Development Program from the Japan Science and Technology. This research is granted by JSPS through the ÅgFunding Program for World-Leading Innovative R\&D on Science and Technology (FIRST Program),Åh initiated by the Council for Science and Technology Policy (CSTP). K.Y. acknowledges the financial support from JSPS. The synchrotron radiation experiment at SPring-8 was done under the approvals of the Japan Synchrotron Radiation Research Institute (2011A1624). The work at KEK-PF was done under the approval of the Program Advisory Committee (proposals 2010S2-001 and 2012G667) at the Institute of Materials Structures Science, KEK.


\begin{thebibliography}{99}
\bibitem{Ref1}
M. Imada, A. Fujimori, and Y. Tokura, Rev. Mod. Phys. \textbf{70}, 1039 (1998). 
\bibitem{Ref2}
Y. Tokura and N. Nagaosa, Science \textbf{288}, 462 (2000). 
\bibitem{Ref3}
 Y. Ogimoto, M. Izumi, A. Sawa, T. Makino, H. Sato, H. Akoh, M. Kawasaki, and Y. Tokura, Jpn. J. Appl. Phys. \textbf{42}, L369 (2003). 
\bibitem{Ref4}
 H. Yamada, M. Kawasaki, and Y. Tokura, Appl. Phys. Lett. \textbf{86}, 192505  (2005). 
\bibitem{Ref5}
T. Harada, I. Ohkubo, M. Lippmaa, Y. Sakurai, Y. Matsumoto, S. Muto, H. Koinuma, and M. Oshima, Adv. Funct. Mater. \textbf{22}, 4471 (2012). 
\bibitem{Ref6}
T. Harada, I. Ohkubo, M. Lippmaa, Y. Matsumoto, M. Sumiya, H. Koinuma, and M. Oshima, Phys. Stat. Sol. RRL \textbf{5}, 34 (2011). 
\bibitem{Ref7}
Note that the lower $T_C$ of our PCMO film than bulk bulk PCMO would be caused by the epitaxial strain from the substrate.
\bibitem{Ref8}
See supporting online materials.
\bibitem{Ref9}
J. Fujihira, A. Uchida, S. Fujihira, M. Furuse, M. Okano, S. Fuchino, K. Yoshimatsu, T. Kadono, A. Fujimori, and T. Koide, J. Cryo. Super. Soc. Jpn.\textbf{48}, 233 (2013).
\bibitem{Ref10}
D. Asakura, T. Koide, S. Yamamoto, K. Tsuchiya, T. Shioya, K. Amemiya, V. R. Singh, T. Kataoka, Y. Yamazaki, Y. Sakamoto, A. Fujimori, T. Taira, and M. Yamamoto, Phys. Rev. B \textbf{82}, 184419 (2010). 
\bibitem{Ref11}
T. Burnus, Z. Hu, H. H. Hsieh, V. L. J. Joly, P. A. Joy, M. W. Haverkort, H. Wu, A. Tanaka, H. -J. Lin, C. T. Chen, and L. H. Tjeng, Phys. Rev. B \textbf{77}, 125124 (2008). 
\bibitem{Ref12}
M. Abbate, F. M. F. de Groot, J. C. Fuggle, A. Fujimori, O. Strebel, F. Lopez, M. Domke, G. Kaindl, G. A. Sawatzky, M. Takano, Y. Takeda, H. Eisaki, and S. Uchida, Phys. Rev. B \textbf{46}, 4511 (1992). 
\bibitem{Ref13}
H. Wadati, A. Maniwa, A. Chikamatsu, I. Ohkubo, H. Kumigashira, M. Oshima, A. Fujimori, M. Lippmaa, M. Kawasaki, and H. Koinuma, Phys. Rev. Lett. \textbf{100}, 026402 (2008). 
\bibitem{Ref14}
S. H\"{u}fner, Photoelectron Spectroscopy (Springer-Verlag, Berlin, 2003).
\bibitem{Ref15}
A. Fujimori, A. Ino, J. Matsuno, T. Yoshida, K. Tanaka, and T. Mizokawa, J. Electron Spectrosc. Relat. Phenom. \textbf{124}, 127 (2002).
\bibitem{Ref16}
A. Ino, T. Mizokawa, A. Fujimori, K. Tamasaku, H. Eisaki, S. Uchida, T. Kimura, T. Sasagawa, and K. Kishio, Phys. Rev. Lett. \textbf{79}, 2101 (1997). 
\bibitem{Ref17}
J. Matsuno, A. Fujimori, Y. Takeda, and M. Takano, Europhys. Lett. \textbf{59}, 252 (2002).
\bibitem{Ref18}
B. T. Thole, P. Carra, F. Sette, and G. van der Laan, Phys. Rev. Lett. \textbf{68}, 1943 (1992). 
\bibitem{Ref19}
T. Koide, H. Miyauchi, J. Okamoto, T. Shidara, T. Sekine, T. Saitoh, A. Fujimori, H. Fukutani, M. Takano, and Y. Takeda, Phys. Rev. Lett. \textbf{87}, 246404 (2001). 
\bibitem{Ref20}
Y. Teramura, A. Tanaka, and T. Jo, J. Phys. Soc. Jpn. \textbf{65}, 1053 (1995). 
\bibitem{Ref21}
Y. Tomioka, A. Asamitsu, H. Kuwahara, Y. Moritomo, and Y. Tokura, Phys. Rev. B \textbf{53}, R1689 (1996).
\bibitem{Ref22}
J. Kanamori, J. Phys. Chem. Solid, \textbf{10}, 87 (1958). 
\bibitem{Ref23}
J. B. Goodenough, Phys. Rev. \textbf{100}, 564 (1955). 
\end{thebibliography}
\end{document}